\documentclass[iop,apj,tighten]{emulateapj}
\usepackage{mathtools, graphicx, booktabs, apjfonts}
\usepackage[usenames,dvipsnames]{color}
 
\def\asc{\lambda}
\def\obliq{\psi}
\def\spinorbit{\varphi}
\def\star{KOI-2138}
\def\planet{\star.01}

\shorttitle{Probable Spin-Orbit Aligned Super-Earth Planet Candidate \planet}
\shortauthors{Barnes, Ahlers, Seubert, \& Relles}

\begin{document}

\title{Probable Spin-Orbit Aligned Super-Earth Planet Candidate \planet}

\author{Jason W. Barnes, Johnathon P. Ahlers, Shayne A. Seubert}
\affil{Department of Physics; University of Idaho; Moscow, ID 83844-0903, USA; ResearcherID:  B-1284-2009}
\email{\tt jwbarnes@uidaho.edu}

\author{Howard M. Relles}
\affil{\tt http://exoplanet-science.com/}

\begin{abstract} 

We use rotational gravity darkening in the disk of \emph{Kepler} star \star~to
show that the orbit of $2.1-R_\oplus$ transiting planet candidate \planet~has a
low projected spin-orbit alignment of $\lambda=1^\circ\pm13$.  \planet~is just
the second super-Earth with a measured spin-orbit alignment after 55 Cancri e,
and the first to be aligned.  With a 23.55-day orbital period, \planet~may
represent the tip of a future iceberg of solar-system-like terrestrial planets
having intermediate periods and low-inclination circular orbits.

\end{abstract}


\section{INTRODUCTION}

Radial velocity and transit surveys have discovered a dramatic variety of
planetary system architectures.  Evidently, the planet formation process need
not necessarily proceed as it has in the solar system.  All of our solar system
planets orbits within $7^\circ$ of the Sun's equatorial plane
\citep{1993ARA&A..31..129L}, an angle that we will call the planets' spin-orbit
alignment angle, $\spinorbit$.  The Sun's planets' spin-orbit alignment
indicates that they formed from a disk without significant subsequent changes in
orbital inclination.

Similarly, measurement of exoplanet spin-orbit alignments can probe those
planets' formation and subsequent orbital evolution.  Determination of the
spin-orbit alignment $\spinorbit$ for Hot Jupiters has provided primary evidence
for ascertaining their origins. Many investigations
\citep[e.g.,][]{2010ApJ...718L.145W,2011ApJ...735...24J,2015arXiv150105565B} 
used ground-based
radial velocity observations of the Rossiter-McLaughlin effect
\citep{1924ApJ....60...15R,1924ApJ....60...22M} (which is sensitive to a
planet's \emph{projected} alignment but not to the orientation of the stellar
spin axis with respect to the plane of the sky) to show that Hot Jupiters around
more massive stars are more likely to be spin-orbit misaligned than are planets
around lower-mass stars.  This difference in alignments probably owes to
evolution, not origins. 

\citet{2012ApJ...757...18A} showed that the transition between mostly-aligned
systems and mostly-misaligned systems occurs near a stellar effective
temperature of $T_\mathrm{\mathrm{eff}}\sim6200~\mathrm{K}$.  Because this is
the border between convective and radiative envelopes for stars,
\citet{2012ApJ...757...18A} postulated that alignment for lower-mass stars
results from tidal interactions.  Later-type, convective stars have higher tidal
dissipation and thus lower tidal quality factors ($Q$) than earlier-type,
radiative stars.  The result is that tides induced on the star by the planet
exchange angular momentum more effectively for low-mass stars.  

Ultimately, this result suggests that Hot Jupiters around low-mass stars are
spin-orbit aligned because the planets pull the stellar spin into alignment over
time via tides --- not because these systems were formed in an aligned state.
The recent discovery of a highly spin-orbit misaligned planet around brand-new
low-mass pre-main-sequence star
PTFO~8-8695~\citep{2012ApJ...755...42V,2013ApJ...774...53B} corroborates the
story that Hot Jupiters acquire orbits isotropically distributed in space
(`random' alignments) early in their history.  

More recent work has used the distribution of measured spin-orbit alignments to
evaluate possible mechanisms to generate spin-orbit misaligned planets in the
first place \citep[see][and references therein]{2014A&A...567A..42C}.  Initial
results indicate that single proposed mechanisms have difficulty reproducing the
observed distribution.  Different systems may therefore produce misalignment in
different ways.  Production mechanisms for the misalignment for smaller,
non-giant worlds have not seen extensive consideration.

Because the Rossiter-McLaughlin effect's signal goes as $\frac{R_p^2}{R_*^2}$
(where $R_p$ and $R_*$ are the planetary and stellar radius respectively),
observations to characterize the origins and evolution of those smaller,
non-giant planets becomes progessively more difficult with decreasing planet
radius.  Some researchers \citep{2012ApJ...756...66H,2014ApJ...796...47M} have
worked around this challenge by constraining the stellar axis tilt with respect
to the plane of the sky (the stellar obliquity $\obliq$) directly by comparing
the stellar radius, rotation period, and projected rotational velocity.  An
obliquity measurement alone constrains but does not directly measure transiting
planets' spin-orbit alignments.  While a non-zero stellar obliquity $\obliq$
requires that any transiting planets be misaligned, a measured obliquity of zero
allows but does not require spin-orbit alignment.  \citet{2014ApJ...796...47M}
thereby indirectly confirmed spin-orbit misalignment for several super-Earths
and super-Earth candidates:  Kepler-96b, KOI269.01, KOI323.01, KOI355.01,
KOI974.01, KOI1890.01, KOI2002.01, KOI2026.01, and KOI2261.01.  Similarly,
asteroseismological determination of a nonzero stellar obliquity $\obliq$ by
\citet{2013ApJ...766..101C} showed that two super-Earths orbiting Kepler-50 and
three super-Earths around Kepler-65 must be misaligned.

Multiple-planet systems can indirectly confirm aligned super-Earths.  If more
than one planet around a given star transits, then the likelihood of planet
coplanarity increases dramatically \citep[e.g.,][]{2011Natur.470...53L}. 
Therefore, Rossiter-McLaughlin measurements of a giant planet can imply a
similar alignment for any other planets in that same system that may be too
small to measure directly.  Hence the Rossiter-McLaughlin determination of
spin-orbit alignment by \citet{2012ApJ...759L..36H} and
\citet{2013ApJ...771...11A} for the giant planet (KOI-94.01) in the KOI-94
system also implies a probable alignment for the 3.73-day period super-Earth in
that system (KOI-94.04).

\begin{figure}[htbp]
\epsscale{1.1}
\plotone{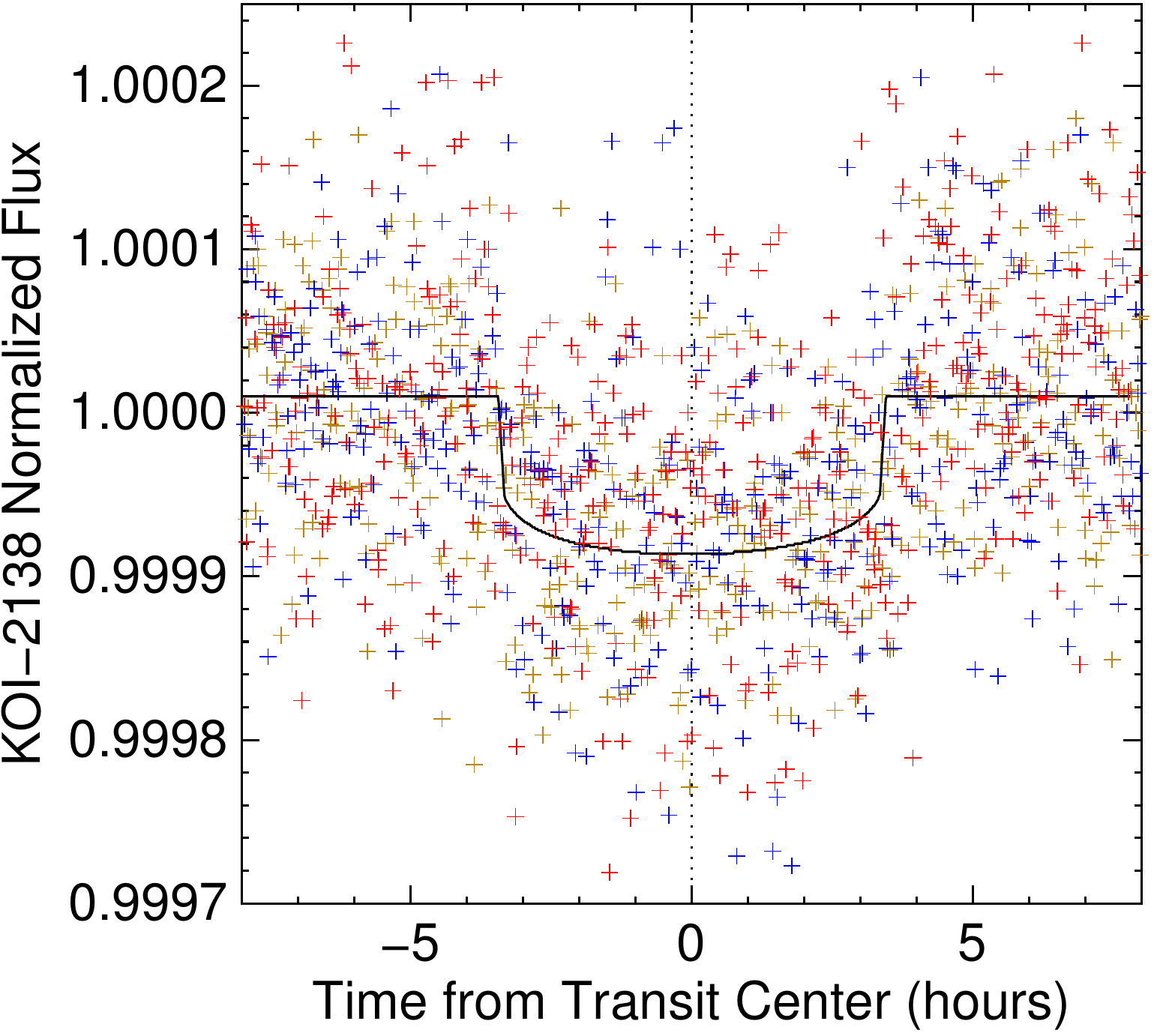}
\caption{\footnotesize  
\label{figure:Qanalysis} This Figure shows the processed \emph{Kepler} data
before binning and as a function of observation Quarter.  Quarter 1 (i.e., all
quarters for which the Quarter number $Q\bmod4=1$) is brown, Quarter 3 is red, and
Quarter 4 is blue.  Quarter 2 is not shown because data from Quarters 6, 10, and
14 were lost due to a CCD failure on-board the spacecraft.  The black line 
indicates best-fit transit lightcurve.}
\end{figure}

The only super-Earth with a directly measured projected spin-orbit alignment to
date is 55 Cancri e \citep[55Cnc e;][]{2014A&A...569A..65B}.  The fourth planet
detected of five now known in the 55Cnc system \citep{2008ApJ...675..790F},
55Cnc e has an orbital period of just 0.7365 days \citep{2010ApJ...722..937D}. 
Its Rossiter-McLaughlin-measured projected misalignment of $72^\circ\pm12$
indicates an askew orbit relative to both the stellar equator and its sibling
planets \citep{2004ApJ...614L..81M}.  The 55Cnc system has clearly experienced a
very different history than that of our own solar system.

The origin and evolution of longer-period (not tidally influenced) planets are
not yet constrained by spin-orbit measurements.  The Rossiter-McLaughlin effect
requires a complete transit to be visible in a given night of ground-based
observing.  Thus Rossiter-McLaughlin measurements for longer period planets are
more difficult because (1) they transit less frequently and (2) their transits
have longer duration.  Additional techniques have successfully measured
spin-orbit alignments of giant planets as well, particularly Doppler tomography
\citep{2010MNRAS.403..151C,2010MNRAS.407..507C,2010A&A...523A..52M,
2012A&A...543L...5G,2014ApJ...790...30J}, stroboscopic starspots
\citep{2011ApJ...733..127S,2011ApJ...743...61S,2011ApJ...740L..10N,
2011ApJS..197...14D,2013MNRAS.428.3671T}, and asteroseismology
\citep{2013Sci...342..331H,2013ApJ...766..101C}.

Herein we use another technique that relies on rapid stellar rotation: 
gravity darkening.  Stellar rotation causes a lower effective surface gravity
($g$) at the equator than at the pole due to centrifugal force.  Lower surface
gravity leads to a larger scale height in the stellar atmosphere, which Von
Zeipel (1924) showed leads to cooler photospheric temperatures.  Those lower
temperatures lead to lower emitted flux from the equator than from the pole,
which we call gravity darkening.  The Von Zeipel Theorem shows that the emitted
flux from a gravity-darkened stellar photosphere is proportional to the
local surface gravity.  Fast-rotating stars therefore have hotter and brighter
poles and cooler and dimmer equators.  Gravity darkening has now been directly
observed by optical interferometric observations of Vega ($\mathrm{\alpha}$
Lyrae)   \citep[][explaining residuals in earlier near-IR interferometry by
\citet{2001ApJ...559.1147C}]{2006Natur.440..896P} and Altair ($\mathrm{\alpha}$
Aquilae)  \citep{2007Sci...317..342M}.  Eclipsing binary stars have long been
analyzed using lightcurves across gravity-darkened stellar disks
\citep[\emph{i.e.}][]{2003A&A...402..667D}.  

\citet{2009ApJ...705..683B} showed that the spin-orbit alignment for planets
orbiting rapidly-rotating stars can be determined from transit photometry alone
by taking advantage of gravity darkening.  The gravity darkening-induced
nonuniformity of the stellar disk introduces characteristic asymmetries into
misaligned planets' transit lightcurves.  Careful fitting of the precise
lightcurve can then constrain an orbiting planet's spin-orbit alignment. 
Because it requires photometry from only a single transit, gravity darkening can
be applied to planets of any orbital period and thus works well for long-period
\emph{Kepler} transits with existing lightcurves.

Gravity darkening has already been leveraged to measure the spin-orbit
misalignment in four systems.  \citet{2011ApJ...736L...4S} first found an
asymmetry in the \emph{Kepler} transit lightcurve of KOI-13 that they attributed
to spin-orbit misalignment around a fast-rotating star --- consistent with the
\citet{2009ApJ...705..683B} predictions.  \citet{2011ApJS..197...10B} then fit
KOI-13.01's asymmetric transit lightcurve with a gravity-darkened stellar
transit model to test whether or not gravity darkening could explain the
measured signature.  It can.  KOI-13.01~has a spin-orbit misalignment of
$56^\circ\pm4^\circ$ \citep{2011ApJS..197...10B}.  The lightcurve analysis in
\citet{2011ApJS..197...10B} is degenerate, so a retrograde spin-orbit alignment
of $126^\circ\pm4^\circ$ is also possible.

Gravity darkening can also indicate spin-orbit alignment in symmetric or nearly
symmetric lightcurves like that of KOI-368
\citep{2013ApJ...776L..35Z,2014ApJ...786..131A}.  In this case, it is the
\emph{lack} of asymmetry in the lightcurve of a planet around a star with
sufficiently high $v~\sin \obliq$ (where $\obliq$ is the stellar obliquity
relative to the plane of the sky, equivalent to the traditional stellar $v~\sin
i$) that constrains alignment \citep[but see also][]{2013ApJ...776L..35Z}.  More
recently, gravity darkened lightcurves revealed nodal precession for exoplanets
KOI-13 \citep{2015Masuda} and PTFO 8-8695
\citep{2012ApJ...755...42V,2013ApJ...774...53B} and mutual alignment for two
transiting planetary candidates (KOI-89.01 and KOI-89.02) that are misaligned
with their parent star (Ahlers \emph{et al.} in preparation).

In this Letter we analyze the \emph{Kepler} transit lightcurve of planet
candidate \planet~to measure its spin-orbit alignment.  At 2.1 Earth radii,
\planet~is only the second super-Earth candidate for which spin-orbit alignment
has been measured, after 55 Cancri e.  In Section \ref{section:observations} we
describe the system's parent star and the \emph{Kepler} photometry from which we
generate a transit lightcurve.  We fit that lightcurve in Section
\ref{section:constraints} before wrapping up with a discussion of the
implications of our measurement in Section \ref{section:discussion}.

\begin{figure*}[tbhp]
\epsscale{1.1}
\plotone{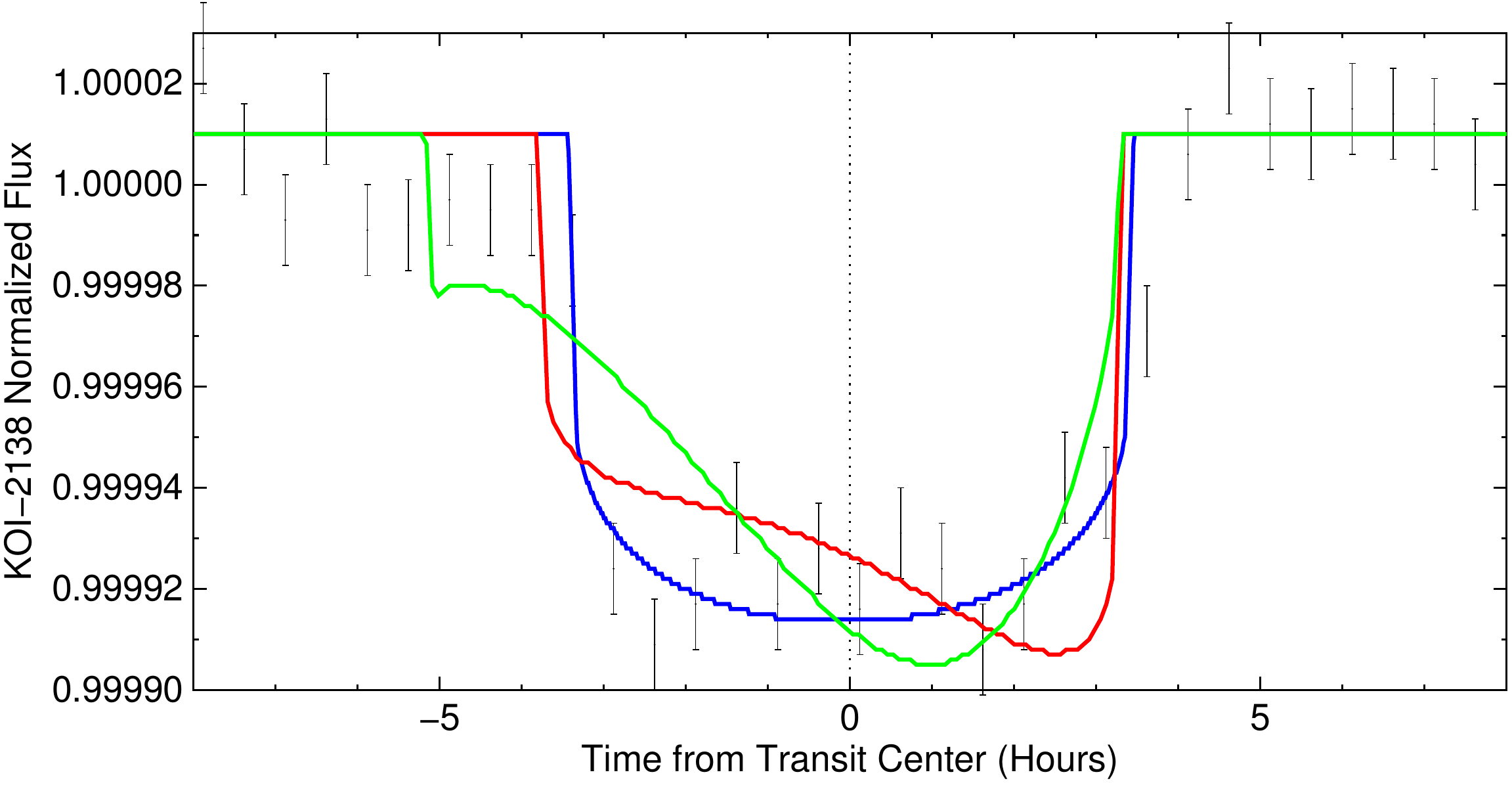}
\caption{\footnotesize Here we show the \emph{Kepler} lightcurve for \star,
centered on the transit and folded with the planet candidate \planet's 23.5541
day period.  The data points in black are binned in time from the original to
aid the in the evaluation of the fits.  The lightcurve of our best-fit
gravity-darkened model is shown in blue, along with two significantly misaligned
models fit to the same data shown in red and green.  The misaligned lightcurves
serve to illustrate the capability of the model to discriminate between aligned
and misaligned transits even with low signal-to-noise.  The jaggedness in the
lightcurves results from the limit of numerical precision of single-precision
floating point numbers, as becomes important for very small transit depths such
as that for \star. 
\label{figure:fitlightcurve}}
\end{figure*}

\section{OBSERVATIONS} \label{section:observations}

The threshold-crossing-event that gave rise to \emph{Kepler} Object of Interest
(KOI) 2138 was first discovered by \citet{2013ApJS..204...24B}.  The most recent
\emph{Kepler} parameters for the parent star show that it is an early-type
$2.335~\mathrm{M_\odot}$ star with $T_\mathrm{eff}=9565~$K
\citep{2015arXiv150107286R}.  The transit depth of the planet is only 86 ppm
($8.6\times10^{-5}$), so although the star has a relatively bright \emph{Kepler}
magnitude of $m_{Kep}=11.98$, the signal-to-noise ratio of the folded transit is
just SNR=30 \citep{2015arXiv150107286R}.

The star's rapid rotation piqued our interest, as fast stellar rotation drives
the gravity darkening effect.  The faster a star rotates, the more gravity
darkened it gets.  On the \emph{Kepler} Community Follow-up Observation Program
(CFOP) site, Allyson Bieryla and colleagues report a spectroscopic
$v~\cos{\obliq}$
for \star~of 200~km/s --- fast enough that the star should be severely
gravity-darkened.  In such high gravity darkening cases the total deviation in
transit depth from a non-rotating star case can be up to a factor of 2
\citep{2009ApJ...705..683B}, and thus eminently visible even in a low
signal-to-noise transit such as this one.

We show the \emph{Kepler} long-cadence photometry of the \planet~transit in
Figure \ref{figure:Qanalysis}.  No \star~data exist for \emph{Kepler} Quarters
6, 10, and 14 due to failure of one of the photometer's detectors.  We use
pre-search data conditioned (PDC) fluxes for our analysis --- the
\planet~transit is subtle enough to be very difficult to identify in raw
photometry.  To eliminate systematic stellar and instrumental variations in the
lightcurve, we median boxcar filter the data with a period of 44 hours ($\sim$6
times the transit duration of 7.1 hours) to remove stellar and instrumental
trends.  As this filter has not previously been used on such a low
signal-to-noise transit, we analyzed its efficacy by generating simulated
lightcurves with similar photon shot noise ($6\times10^{-5}$) and
Gaussian-distributed random linear trends with total differences of $10^{-4}$
over the 44-hour filter period.  When we process these simulated data in the
same way as we do the \emph{Kepler} PDC data, the resulting processed
lightcurves retained any original symmetry or asymmetry of a small transit of
the same depth as that of \planet.  The median boxcar cannot account for
higher-order variability on timescales shorter than the boxcar duration,
however, and we rely on phase-folding of the data to average out any variations
on those shorter timescales.

No trends or variations are evident when the timeseries are analyzed as a
function of \emph{Kepler} viewing geometry (i.e. looking at Quarters 1, 5, 9,
13, and 17 together because all were acquired with the same spacecraft
orientation and similar for the Quarter 2, 3, and 4 geometry; see Figure
\ref{figure:Qanalysis}).  The \star transit lightcurve shows no evidence for
significant transit timing variations, either --- the strongest periodicity in a
Lomb-Scargle periodogram of the $O-C$ times occurs with a false alarm
probability of 0.269 at 49.93 days.  Therefore we folded the 44 transits with
data at the 23.5540725-day MAST-listed period.  Because experience indicates
that only systems with false alarm probabilities less than 0.01 possess credible
transit timing variations, our folding the lightcurve at the observed period
does not adversely affect the resulting analysis.  And folding the data improves
the signal-to-noise of the final lightcurve while averaging out variable
systematic influences as may arise from PDC conditioning, the median boxcar
filter step, or inherent stellar variability.

\renewcommand{\arraystretch}{1.5}
\begin{table}[!htbp]
\centering
\begin{tabular}{|c|c|}
\hline
Parameter & Best Fit Values \\ \hline
$\chi^2_\mathrm{reduced}$ & 1.0424 \\ \hline
$R_*$ & $2.286 \mathrm{R_\odot}$ (fixed)\\ \hline
$M_*$ & $2.335 \mathrm{M_\odot}$ (fixed)\\ \hline
$V\sin\obliq$ & $200 \mathrm{\frac{km}{s}}$ (fixed)\\ \hline
$\beta$ & 0.25 (fixed) \\ \hline
$R_\mathrm{p}$ & $2.1\pm0.4~\mathrm{R_\oplus}$ (fit) \\ \hline
$i$ & $88.34^\circ\pm0.11^\circ$ (fit) \\ \hline
$c_1$ & 0.49 (fixed) \\ \hline
$c_2$ & 0 (fixed) \\ \hline
$e$ & 0 (fixed) \\ \hline
$\asc$ & $1^\circ\pm13^\circ$ (fit) \\ \hline
$\obliq$ & $-4^\circ \pm60^\circ$ (fit) \\ \hline 
$P_\mathrm{rot}$ & $14$ hr (derived)\\ \hline
$f_*$ & 0.10 (derived)\\ \hline 
\end{tabular}
\caption[justification=justified]{\label{table:parameters} Best-fit parameters for the \star~system.  
The indicated value for the projected alignment $\asc$ represents the formal
$1-\sigma$ uncertainty from the covariance matrix --- a more complete analysis
allows for $\asc=1^\circ~^{+50}_{-20}$, albeit with improbably
lucky transit geometries (see text).}
\end{table}

\section{CONSTRAINTS}\label{section:constraints}

To constrain the spin-orbit alignment $\spinorbit$ of planet candidate \planet,
we fit the lightcurve using the \cite{2009ApJ...705..683B} {\tt transitfitter}
algorithm.  This program numerically integrates flux from gravity-darkened stars
and fits the resulting lightcurves using a Leavenberg-Marquardt approach from
\citet{NumericalRecipes} with time-integration.  Although the signal-to-noise
ratio of the transit is low \citep[29.9;][]{2015arXiv150107286R}, the large
deviations from symmetry expected for a misaligned system (e.g., the red and
green curves in Figure \ref{figure:fitlightcurve}) allow us to rule out some
projected alignments $\asc$.

\begin{figure}[htbp]
\epsscale{1.1}
\plotone{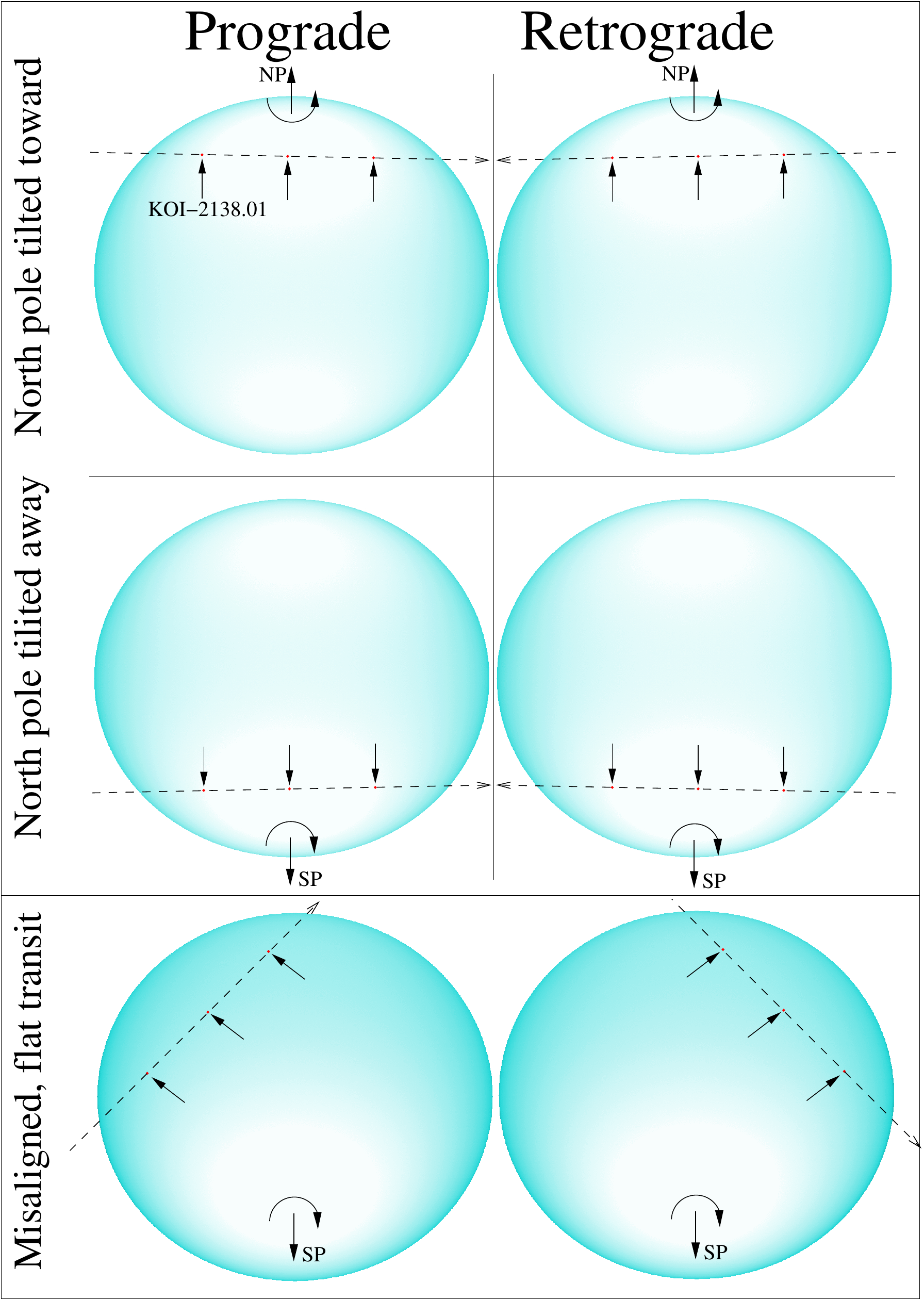}
\caption{\footnotesize  
\label{figure:angles} Our allowed transit geometries for \planet.  With small
super-Earth candidate \planet~orbiting a relatively large, early-type star, the planet as
shown here is tiny.  We denote the planet's position with arrows and show it in red to
make it more evident.}
\end{figure}

The small planet radius leads to rapid transit ingress and egress, unresolved by
\emph{Kepler} long-cadence photometry.  Hence in our fits we fix the stellar
radius at the MAST value of $2.286~\mathrm{R_\odot}$.  The low signal-to-noise
cannot independently constrain the stellar limb darkening; therefore we fix the
$c_1$ ($=u_1+u_2$) limb darkening coefficient at the value measured for KOI-13
\citep{2011ApJS..197...10B}, which is at a similar $T_\mathrm{eff}$ \citep[but
see also][]{2015Masuda}.  We hold the stellar $v~\cos \obliq$ fixed at 200~km/s,
which sets the intensity of stellar gravity darkening.  \citep[We also assume
that the gravity darkening parameter $\beta=0.25$ as expected for stars with
radiative envelopes][]{1924MNRAS..84..665V}.  However, the $v~\cos \obliq=$~200~km/s
measurement is a lower limit to the star's rotational velocity because it
represents the highest valid $v~\cos \obliq$ from the template spectra (D. Latham,
personal communication); thus more stringent constraints may be possible with better
stellar $v~\cos \obliq$ determination.

We show the best-fit lightcurve in Figure
\ref{figure:fitlightcurve}, and the best-fit parameters in Table
\ref{table:parameters}.  We depict the transit model graphically in Figure
\ref{figure:angles}.  The stellar obliquity $\obliq$, defined as how far the
star's north pole is tilted away from the plane of the sky, is only very poorly
constrained ($\obliq=-4^\circ \pm60^\circ$).  In fact a first-principles
calculation shows a similar constraint:  if $\obliq$ were beyond $\pm60^\circ$
the star would be rotating beyond its breakup speed with $v~\cos \obliq =
200~\mathrm{km/s}$.  Without a more robust estimation of the stellar obliquity
the true spin-orbit alignment of this system will remain unknown.

We derive more useful constraints on the projected alignment $\asc$, defined as
the direction of the planet's velocity vector at inferior conjunction measured
clockwise from the $x$-axis (to the right in Figure
\ref{figure:angles}).  The formal uncertainty from the fit covariance matrix
yields $\asc=1^\circ\pm13^\circ$ --- consistent with \planet~in spin-orbit
alignment.  

A more thorough error analysis shows, however, that valid models with $\asc$ up
to $60^\circ$ or as low as $-20^\circ$ also exist.  When we explore error space
by fixing $\asc$ and fitting for the remaining parameters \citep[page
815]{NumericalRecipes}, we find that specific combinations of parameters can
replicate symmetric, flat-bottomed transits with spin-orbit misaligned planets. 
Those models contrive to have the planet traverse specific stellar chords that
have nearly uniform flux due to combinations of gravity darkening and limb
darkening, as shown in Figure \ref{figure:angles} at bottom.  While we cannot
rule such fortuitous transit chords out, we consider them to be less probable
than the spin-orbit aligned variants that show stronger robustness to the
parameters that we held constant (i.e. $R_*$, $e$).  Independent measurements of
the stellar obliquity $\obliq$, such as with asteroseismology, could resolve the
degeneracy.

\begin{figure}[htbp]
\epsscale{1.1}
\plotone{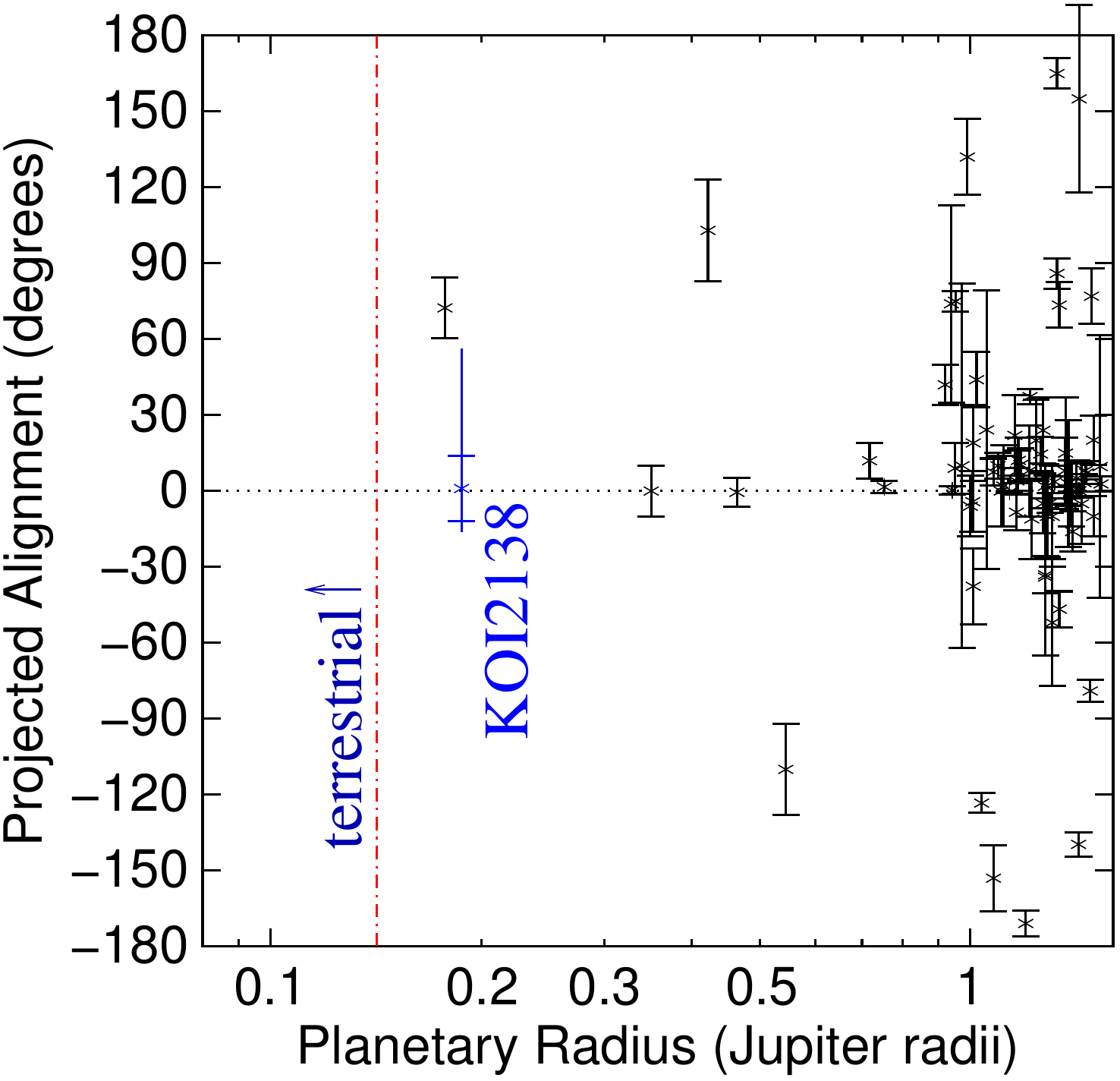}
\caption{\footnotesize Here we plot the measured projected spin-orbit alignment
$\asc$ as a function of planetary radius $R_p$ in Jupiter radii
$R_\mathrm{Jup}$.  Each exoplanet with a useful measured constraint is shown on
this plot, with data from Ren\'e Heller's
Holt-Rossiter-McLaughlin Encyclopaedia ({\tt
www.physics.mcmaster.ca/\textasciitilde rheller}).  The red vertical line indicates the
$1.6~R_\oplus$ cutoff above which all planets are volatile-rich according to
\citet{2014arXiv1407.4457R}.  \star, from this work, is indicated in blue and
is the first spin-orbit aligned super-Earth candidate.
\label{figure:heller_alignments}}
\end{figure}

\section{DISCUSSION}\label{section:discussion}

\planet~represents just the second super-Earth candidate with a measured spin-orbit
alignment and the first to be (probably) aligned.  In Figure
\ref{figure:heller_alignments} we show the projected spin-orbit alignment $\asc$
for planets as a function of their radius.  While we now have an appreciable
understanding of the spin-orbit alignments of gas giants in short (less than 10
days) orbits, only a few planets smaller than Saturn have had their spin-orbit
alignment determined.  

The only other $\asc$-measured super-Earth, 55Cnc e, has a 0.74-day misaligned
orbit.  Therefore if it is terrestrial, consistent with its Earth-like density,
then 55Cnc e must be very different from any solar system planet.  Certainly its
close orbit drives a very high equilibrium temperature of over 1600K
\citep{2011ApJ...740...49V}.  Furthermore,  that orbit is in fact so close-in
that 55Cnc e almost certainly did not form \emph{in situ}, and indeed its
spin-orbit misalignment implies an interesting dynamical history as well.

Constrastingly, \planet's longer-period (23.55 day), aligned orbit indicates that
it could potentially be the first representative of an expected population of
solar-system-like terrestrial planets.  Its semimajor axis of $0.21$~AU puts
\planet~inward of the Sun's Mercury by a factor of two, and with its hotter star
\planet~should still be hot with a subsolar $T_\mathrm{eff}$ of $\sim1300$K. 
Terrestrial planet formation at such distances may be difficult, but given our
lack of knowledge of planet formation around early-type stars this problem may
not be insurmountable.  If \planet~did indeed form near its present location,
then its aligned orbit may portend of a large population of rocky planets that
we may be able to characterize in the coming decades.

\acknowledgements  The authors thank the anonymous reviewer for constructive
comments.  The authors acknowledge support from the NASA ADAP Program, grant
\#NNX14AI67G.  This study has made use of Ren\'e Heller's
Holt-Rossiter-McLaughlin Encyclopaedia ({\tt
www.physics.mcmaster.ca/\textasciitilde rheller}).


\end{document}